# Substrate Mediated Synthesis of Ti-Si-N Nano-and-Micro Structures for Optoelectronic Applications

*Sachin Yadav, Alka Sharma, Bikash Gajar, Mandeep Kaur, Dinesh Singh, Sandeep Singh, Kamlesh Kumar Maurya, Sudhir Husale, Vijay Narain Ojha and Sangeeta Sahoo**

S. Yadav, Dr. A. Sharma, B. Gajar, Dr. M. Kaur, Dr. S. Husale, Dr. V. N. Ojha, Dr. S. Sahoo
Time & Frequency and Electrical & Electronics Metrology Division, CSIR-National Physical Laboratory, Dr. K. S. Krishnan Marg, New Delhi, India- 110012
E-mail: sahoos@nplindia.org
S. Yadav, Dr. A. Sharma, B. Gajar, Dr. K. K. Maurya, Dr. S. Husale, Dr. V. N. Ojha, Dr. S. Sahoo
Academy of Scientific and Innovative Research (AcSIR), CSIR-National Physical Laboratory, Dr. K. S. Krishnan Marg, New Delhi, India- 110012
D. Singh, S. Singh, Dr. K. K. Maurya
Sophisticated Analytical Equipments, CSIR-National Physical Laboratory, Dr. K. S. Krishnan Marg, New Delhi, India- 110012

Keywords: ternary silicide & nitrides, self-assemblies, surface plasmon polaritons, titanium nitride, Drude's resistivity

**Abstract**

Being one of the strongest materials, ternary TiSiN exhibits a very interesting family of binary transition metal nitride and silicide systems. A novel technique to fabricate morphologically fascinating nano and micro structures of TiSiN is reported here. The referred TiSiN films, majorly constituted with cubic TiN phase, are enriched with crystalline nanoparticles, micro-flowers and faceted micro-crystals which possess attractive functionalities towards plasmon mediated optoelectronic applications. Reactivity of titanium to silicon nitride based dielectric topping on the substrate at high temperature plays the key role in nitride formation for the demonstrated protocol. The optoelectronic response for these morphologically enriched composite films indicates an influential role of photo-induced surface plasmon polaritons (SPPs) on their dc transport properties. A plasmonically tuned resistive switching, controlled by the surface morphology in association with the film thickness, is observed under light illumination. Using Drude's modified frequency dependent bulk electron scattering rates and surface mediated SPPs-electron scattering rates, a generic model is proposed for addressing unambiguously the increased device resistance in response to light. The featured synthesis process opens a new direction towards the growth of transition metal nitrides while the proposed model serves as a basic platform to understand photo-induced electron scattering mechanisms in metal.

## 1. Introduction

In the search for low loss plasmonic materials for the realization in useful device applications, transition metal nitrides (TMN) have garnered significant attention over the conventional metals like gold and silver due to their compatibility to CMOS technology, high thermal and mechanical stability and their ceramic nature.[1] Among the TMN series, TiN and ZrN are the most popular candidates for plasmonic studies.[2] However to have better performance with reduced optical loss, a material should have reduced carrier density and metallic properties.[1,3] In this regard, a new class known as conducting ternary TMN have shown great promises by appropriate doping of a binary TMN with a third element.[4] For example, TiZrN is expected to have better plasmonic properties and performances compared to its binary constituents TiN and ZrN.[1] Further, recently ZrTaN, TiTaN, and TiAlN are shown to have interesting plasmonic properties.[4] In the class of ternary TMNs, one of the corrosion resistant hardest materials is Ti-Si-N,[5] which exhibits an excellent chemical and thermal stability at high temperature and possesses extra ordinary mechanical performances[5,6] along with interesting interfacial properties.[7] As Ti-Si-N can be considered as Si doped TiN with the later as emerging high performance refractory plasmonic metal,[8-10] it can be considered as a promising material for plasmonic applications but best of our knowledge Ti-Si-N has not been explored as plasmonic and/or photonic material.

In addition to materials selection, the surface morphological variations in reduced dimension add extra functionality to the material's performance. For example, spherical nanoparticles,[11] nanowires,[12,13] nanoprism and antennas,[14,15] faceted crystals,[16] nano stars,[17-19] and flowers,[20,21] self-assembled ordered crystalline superstructures[22,23] are among the popular morphologies for optoelectronic, photonic and plasmonic applications.[24] However for Ti-Si-N, a very little attention has been paid to the new growth mechanism, morphology controlled evolution of nanostructures and their optoelectronic / plasmonic applications.

Therefore, it is very important and necessary to develop a cost effective, reliable and simple synthesis protocol which can offer interesting morphological variations for adding new functionalities towards better performance. Here, we demonstrate that by using high vacuum annealing of Si/Ti/Si based thin film tri-layers on $Si_3N_4$/Si (100) substrate, one can successfully fabricate crystalline nanoparticles, faceted micro-crystals, and crystalline micro-flowers on Ti-Si-N based nanostructured composite film (NCF) which are predominantly occupied with cubic fcc-phased nano and micro structures of TiN,[1] Coupling with optical excitation in the visible and NIR wavelength range, the Ti-Si-N based NCFs having devices on top exhibit surface morphology dependent anomalous photo-response which shows switching to a highly resistive state.

Recently, there are many reports of negative photoconductivity (NPC) observed in materials like semiconducting nanostructures,[25-27] topological insulators,[28] graphene based composite systems,[29-31] metallic nanostructures[32-35] etc. Trapping of photo-excited electrons by localized trapped states,[36] band gap opening,[28] e-SPPs scattering mechanisms[32] are commonly used to address the origin of the NPC. However, if a generic model were available to understand the origin of NPC particularly for metallic systems, the performance could be tuned in a preferable way by controlling the morphology. Considering the possible scattering mechanisms for the conduction electrons, we introduce a simple but competent model primarily constituted with frequency dependent bulk electron scattering rate and surface mediated scattering of electrons with photo-induced SPPs[32]. The observed NPC, controlled by the morphology induced SPPs and the film thickness, can be justified and explained unambiguously by the presented generic model.

Conventionally, Ti-Si-N based NCFs are grown either using reactive dc sputtering of Ti & Si or a combination of other Ti-Si phases in presence of Ar/$N_2$ gas mixtures[6,37,38] or by high temperature reactions of TiN with Si.[7] Here using the established knowledge of high temperature reactivity of Ti to $Si_3N_4$,[7,39,40] we report a novel technique which employs $Si_3N_4$/Si substrate as the only source of nitrogen in the presented Ti-Si-N based NCFs growth protocol. As above 950 °C $Si_3N_4$ spontaneously decomposes into $N_2$ and Si, formation of TiN and various Ti-Si phases can be achieved during the interaction between Ti and $Si_3N_4$.[39] The present work reports annealing temperature of ~ 800˚C and a thin amorphous Si (*a-Si*) layer as a barrier between Ti with $Si_3N_4$ dielectric on the substrate in order to avoid their direct contact for favourable growth of Ti-Si-N based NCFs.[39] Further, we use an additional *a-Si* top capping layer of same thickness on Ti to protect it from the adsorbent contaminants present in the ambient.

## 2. Experimental Section

We employed highly doped Si (100) wafer covered with LPCVD grown $Si_3N_4$ layer of ~ 40 nm thickness as the starting substrate. The substrates were cleaned by consecutive sonication in acetone and isopropanol bath for 15 minutes each, followed by oxygen plasma cleaning of 10 minutes. Cleaned samples were loaded into the sputtering chamber where we heated the substrates at ~ 800 ˚C for 30 minutes in very high vacuum conditions ($p$ ~ 6.6 x $10^{-6}$ Pa) to remove any adsorbed or trapped molecules/residue on the surface of the substrate. Once the samples were at room temperature we sputter coated them with an assembly of *a-Si* (~ 8 nm)/Ti (15-40 nm/*a-Si* (~ 8 nm) tri-layer. Thickness variation was done by rotating the sample holder with respect to an axis parallel to the target plane. Sputtering of Si (99.995% purity) was performed with a base pressure less than 3.5 x $10^{-9}$ Torr in the UHV chamber while Ti (99.995% purity) was sputtered in the adjacent loadlock chamber with a base pressure less than 1.5 x $10^{-7}$ Torr. Finally, the sputtered samples were annealed at ~ 800 ˚C for 2 hours at pressure ~ 5 x $10^{-8}$ Torr. The whole process is presented schematically in **Figure 1**(a)-(d). Morphological characterization of the samples were performed by using FESEM (Zeiss) and AFM (Multimode V, NS V, Veeco) in tapping mode, whereas structural characterizations were performed by HRTEM (FEI, Technai, 300 KV, G2, STWIN) with point resolution 0.205 nm & line resolution 0.144 nm. Grazing incidence X-ray diffraction (GIXRD) characterizations were done by PAnalytical PRO MRD X'pert X-ray diffractometer with $CuK_\alpha$ radiation operating at 40KV and 30 mA. Elemental compositional analyses on different shaped microcrystalline structures were done by EDS attached with FESEM.

Photoconductivity measurements were performed at room temperature in a probe-station from Cascade Microtech with shield enclosure using Keithley 2634b as the source measure unit. The light sources are white light from halogen lamp and 532 nm laser in the visible part and 1064 nm laser in the NIR region.

**3. Results and Discussion**

We have defined three different zones having variations in Ti thickness while keeping the top and bottom *a-Si* layers fixed. Zone-I, zone-II and zone-III correspond to the thinnest Ti layer (thickness ~ 15 nm), an intermediate Ti layer (thickness ~ 25 nm) and the thickest Ti layer (thickness ~ 40 nm), respectively. The whole synthesis process is presented schematically in Figure 1(a)-(d) with Figure 1(b) presenting the three zones. The overall morphological changes, as observed by field emission scanning electron microscopy (FESEM) images for the referred three zones, are presented in Figure 1(e)-(g). Zone-I [Figure 1(g)] with the thinnest Ti layer (15 nm) produces a densely packed uniform mat consisting of nanoparticles with dimensions in the range of 30-40 nm. With a gradual increase in Ti thickness up to about 25 nm for zone-II [Figure 1(f)], the granular nature for the surface becomes clear with grain sizes in the similar range as that emerged for zone-I. It is apparent that zone-III [Figure 1(e)] with the thickest Ti layer (~40 nm) is morphologically very different than the other 2 zones. There are diversities in the size and shape of the structures appeared on the surface of the film. Hence, thickness of Ti plays a crucial role for the morphologically evolution.

In **Figure 2**(a)-(s), we have presented a collection of FESEM images from zone-III displaying a variety of structures grown successfully using the prescribed method. The distinctly visible structures as presented in Figure 1(e) are actually surrounded by a non-uniform floating layer, resembling of scattered clouds constituted with small granular particles of 60-70 nm dimension, as shown in Figure 2(a). These floating particles are most likely the initial stages of the bigger structures and may act as the seeds which self-assemble to form 2D and/or 3D crystalline Ti-Si-N microstructures. We have collected the relatively bigger structures from zone-III and have displayed them categorically in Figure 2(b)-(s). The FESEM micrographs for the first set of representative faceted microcrystals with sharp crystalline planes in diverse shapes and sizes are shown in Figure 2(b)-(i). The bottom two rows present the formation of the flower-like structures. The second last row [Figure 2(j)-(n)] displays the structures which are one step prior to the formation of complete round-shaped flowers while the bottom row [Figure 2(o)-(s)] displays the completely rounded micro-flowers. Images in Figure 2(m) & (n) clearly demonstrate the layering of crystalline planes to resemble flower-like crystal structures while Figure 2(l) indicates how the different petals are assembled at the center to complete the formation of a branched rounded flower. Micrographs in (j) & (k) of Figure 2 show the partially completed assembling of the micro flowers while the bottom row displays varieties of flower like round shaped closed (o) and/or branched (q) structures formed by self-assembly and layering of the crystallites. It should be noted that all the presented structures from zone-III in Figure 2 can be found in a single sample. However, we have selectively categorized them into several growth sequences/stages in order to understand the total growth and assembling process. The nanoparticles, shown in Figure 2(a), self-assemble sequentially first into small crystals followed by two dimensional crystalline flakes and/or petals and finally to the big faceted micro crystals [Figure 2(e)-(i)] and micro-flowers (bottom two rows). The whole self-assembly process for the 3-D microstructures occurs by the layered growth of the 2D crystalline flakes/petals during the high vacuum annealing at 800 ˚C for 2 hours for the zone-III samples.

The structural characterizations of the as-grown Ti-Si-N based NCFs are performed by x-ray diffraction (XRD) and transmission electron microscopy (TEM). **Figure 3**(a) presents the XRD pattern obtained from zone-III for Ti-Si-N NCFs with Ti thickness ~40 nm. The prominent peaks confirm the formation of cubic TiN phase along with TiSi phases. A strong peak related to hexagonal Ti (102) is also present indicating the presence of excess Ti. A couple of weak peaks are present which we denote as the unknown peaks but they might correspond to other Ti-Si phases. Note that the strong peak observed at $2\theta = 36.8^\circ$ is closely related to TiN (111) and TiSi (210) planes. For clarity, we present an enlarged view of the dotted rectangular portion separately in the inset of Figure 3(a) which presents two distinct peaks related to TiN (111) and TiSi (210) phases. In addition, the presence of TiSi (211) plane along with other strong TiN peaks indicates the co-existence of both Ti-N and Ti-Si phases in the Ti-Si-N composite film. For further confirmation of the presence of the silicide phases, we have performed XRD characterization of a controlled sample grown at the same batch as that shown in Figure 3(a) on a different substrate ($SiO_2$/Si) and the comparison in XRD spectra is presented in (Figure S1 in (Supporting Information) . Clear indication of various titanium silicide formation is evident in the XRD data for the control sample in Figure S1(b). However, it is evident that the Ti-Si-N based NCFs are predominantly occupied with cubic TiN phase as all the strong peaks of TiN appeared in the XRD pattern. Here, the TiN (111) peak appears to be stronger than the (200) peak and this might be due to the presence of both TiN (111) and TiSi (210) planes almost at the same $2\theta$ position. This is noteworthy to mention that the plane (210) corresponds to the strongest peak for orthorhombic TiSi.

For further insight into the structural properties, we have performed TEM analysis along with selected area electron diffraction (SAED) studies. The bright field TEM micrographs of a randomly selected region from the granular mat taken from zone-III are presented in Figure 3(b)-(g). The scattered particles/grains having dimensions in the range of ~ 20-30 nm appear as the dark quasi-circular spots. The corresponding SAED pattern is shown in Figure 3(c). The circular ring type of patterns appeared in SAED indicate the polycrystalline nature of the film. In the reciprocal space the indexed rings indicate the presence of fcc-TiN phases for (111), (220), (311) planes all of which appear in the XRD pattern too. The diffraction spots constituting the inner circle are strong and wide indicating a possible coincidence with TiSi (210) planes as it was assumed for the XRD pattern shown in Figure 3(a). The selected areas, encircled with white dotted circles with uppercase roman numbers in Figure 3(b), are elaborated with atomic scale images acquired by high resolution TEM (HRTEM) in Figure 3(d)-(g). HRTEM images show the crystalline nature of the particles. The closest possible indexing reveal and confirm the presence of (200) and (111) planes from TiN along with several Ti-Si phases, *viz.* $TiSi_2$ [(311), (020)], TiSi [(210), (301)] etc. Therefore, the HRTEM clearly indicates the presence of TiN, TiSi and $TiSi_2$ phases in the fabricated Ti-Si-N based NCFs. We have studied many places from several batches of samples and have found similar results.

Following the granular base mat we have studied individual microstructures such as flakes and flowers under TEM. **Figure 4** displays HRTEM images along with the elemental analysis performed using energy dispersive spectroscopy (EDS) for a micro flower [Figure 4(a)-(d)] and a microcrystal flake [Figure 4(e)-(i)]. Full EDS spectra for the flower and microcrystal flake are shown in Figure S2. The FESEM image of the flower is presented in Figure 4(a).The flower looks differently than those presented in Figure 2 and this could be just because as-grown flowers are attached to the substrate rigidly with their branches stretched whereas the plucked flower placed on the TEM grid possesses rather freely floating branches. The EDS based elemental analysis shown in Figure 4(b) confirms the presence of all three elements in the displayed Ti-Si-N flower. A TEM micrograph of the same is presented in Figure 4(c) and a magnified image of a selected part is shown in Figure 4(d). The layer by layer assembling of 2D crystalline planes/sheets of different shapes to form the petals and subsequently the complete flower is evident in Figure 4(d).

Similarly, a microcrystalline flake was characterized by using TEM and EDS [Figure 4(e)-(j)]. The TEM micrograph of the flake is shown in Figure 4(e). A relatively thin part, shown by the white dotted circular region in Figure 4(e), is studied by HRTEM and the possible crystalline phases present in the flake are shown in Figure 4(f)-(i). The (200) orientation of TiN is present here too as that was evident in XRD and in HRTEM results on the mat shown in Figure 3. Additionally, we observe here the presence of $Ti_5Si_3$ phase. The EDS elemental analysis of the flake is shown in Figure 4(j). The presence of all the three elements in the flake is confirmed. The structural analysis confirms a dominant presence of the binary constituent TiN in Ti-Si-N and the morphological studies exhibit the diversity in the formation of nano and microstructures with different shapes and sizes interesting for optoelectronic and plasmonic applications. Here we have fabricated two terminal electrical devices with Au (100 nm)/Ti (5 nm) contact leads on the films from different zones [**Figure 5**(a) and (b)]. The structures like rods, petals, and self-assembled flower like structures floating on the granular base mat are clearly evident in Figure 5(b). The linear dark state *I-V* characteristic, shown in the inset of Figure 5(b), indicates the formation of Ohmic contact at the interface between the leads and the film. Time resolved photocurrent ($I_{ph}$) defined as, $I_{ph}=I_{light} - I_{dark}$, measurements in visible and NIR spectral range for zone-III sample are shown in Figure 5(c). For all the three light sources (halogen, 532nm and 1064 nm), the source-drain current ($I_{sd}$), with a fixed $V_{sd}=700\ mV$, strongly responds to light illumination, i.e. the devices switch to a higher resistive state in response with light as it is observed in many recent reports.[32] Further, we have observed a linear dependence of $I_{ph}$ with the bias voltage [Figure S3(a)&(b)] and laser power for NIR light [Figure S3(c)&(d)].

### 3.1. Morphology Controlled Photo-response

To have a primary understanding over the role of the micro and nano structures, *viz.* nano-grains, microcrystals, micro-flowers, and the densely packed thin film mat on responding the light illumination to their dc electrical conductivities, we have measured time resolved photocurrent for three different samples chosen from each of the mentioned three different zones. Representative AFM images, having 10 μm x 10 μm dimensions for zone-III, zone-II & zone-I, are displayed in **Figure 6**(a), (c) & (e), respectively. Figure 6(a) for zone-III clearly displays the structures like nanograins, microcrystalline flowers, rods etc. For clarity, nanograins, bounded by the white square in Figure 6(a), are highlighted in magnified image presented in Figure 6(b). The grain sizes for these particles are in the range of 60-70 nm. High resolution images bounded by an area of 1 μm x 1 μm for zone-II & zone-I are shown in Figure 6(d) & (f), respectively. Few differences appear for the surface morphology and the grain sizes for zone-II and zone-I samples. Thickness variations, shown by the color scale, for the surface structures related to the studied three zones are evident. However, the lateral sizes of the grains on zone-II samples are in the range of 20-30 nm whereas the same for zone-I samples appear to be in the range between 30 to 40 nm. In Fig. 6(g), we present the related photoresponse for the white light illumination from a halogen lamp. For a comparison, the time dependent source-drain current for a fixed bias, $V_{sd}$ = 600 mV, with two consecutive light 'OFF' and 'ON' cycles for zone-III, zone-II, and zone-I are exhibited in top, middle, and bottom panels, respectively. First, we observe that the overall source-drain current decreases with decreasing thickness as thicker samples are expected to be less resistive compared to the thinner ones. Second, the amplitude of $I_{ph}$ [top panel of Figure 6(g)] is more for the thickest sample [zone-III] embedded with nanograins, microcrystals and microflowers, than that for the zone-II samples with no extra structures on the surface of the granular base mat. Finally, it is very interesting that there is almost no detectable photo-response in the measured $I_{sd}$ [bottom panel of Figure 6(g)] for the densely packed base mat with the thinnest film from zone-I. As the surface structures look similar for zone-II and zone-I samples, the thickness variation may play the dominant role for the observed anomaly in the photo-response for these two zones. Thus it is obvious from Figure 6 that the light effect on the dc conductivity strongly depends on the surface morphology and the film thickness. Note, for providing maximum response we present selectively the morphology dependent photocurrent only for halogen light. With 532 nm laser we observe approximately similar effect (Figure S4) while only zone-III samples respond strongly to1064 nm laser [Figure 5 & Fig. S3(c)-(d)] and no significant change in $I_{sd}$ is observed for zone II & zone-I samples with 1064 nm laser.

### 3.2. Mechanism behind photo-response: resistivity calculation and the influence of light on the dc resistivity

In order to understand the origin of the observed anomalous photoconductivity controlled mainly by the surface morphology and the film thickness, here we calculate the dc resistivity in the dark and illumination conditions.

As the devices exhibit linear current-voltage characteristics [inset of Figure 5(b)] the device current under illumination ($I_{light}$) and the change in current due to the light excitation ($I_{ph}$/ $\delta I$) can be expressed as:

$$I_{light} \propto \frac{1}{\rho_{dark}}\left(1 - \frac{\delta\rho}{\rho_{dark}}\right) \quad (1)$$

and

$$I_{ph} = \delta I \propto \frac{\delta\rho}{\rho_{dark}^2} \quad (2)$$

Where, $I$ and $\rho$ represent current and resistivity, respectively. The subscript *dark* (*light*) represents the measured values under light '*OFF*' ('*ON*') state, respectively. $I_{ph}$ is the change in current due to photon excitation.

From Equation (2) $I_{ph}$ depends on dark state resistivity ($\rho_{dark}$) and the photo-induced resistivity ($\delta\rho$). Thus for a clear understanding of the origin behind the observed NPC, we now consider the effect of surface morphology, thickness and the light illumination on the device resistivity. In general, $\rho_{dark}$ is constant for a device when only the effect of light is considered while keeping the device parameters like the film thickness, surface roughness etc. unaltered. In this case, $I_{ph}$ depends only on $\delta\rho$. In order to calculate the dc resistivity, $\rho$, we consider Drude's free electron model as, $\rho_{dc} = \frac{m_e^*}{n_e e^2 \tau} \propto \frac{1}{\tau}$, where, $n_e$, $m_e^*$, $e$ are the free electron density, effective mass, and electronic charge, respectively. $\tau$ represents the electron scattering time which depends mainly on the purity of the metal, charge state of the impurity and temperature. Considering $m_e^*$ and $n_e$ as constants in the dark state, the dc resistivity depends mainly on the scattering time, $\tau$. Electron-phonon, electron-impurity and electron-electron scattering mechanisms are the dominant parts towards the bulk electron scattering in a metal. Surface electron scattering, controlled mainly by the surface roughness, grain size and grain boundary, also contribute significantly to the resistivity.[41]

Under the light illumination, surface plasmons (SPs), initiated and controlled by the surface roughness and morphology, takes a dominant role to control the device's transport response. Surface morphology controls the local electric field enhancement,[21,42,43] radiative damping,[44] and non-radiative Landau damping[45,46] for the SPs.[47] Further, coupled with the light excitation SPs can induce SPPs which contributes significantly to the total electron scattering rate while interacting with the conduction electron.[32] The electron-SPPs scattering rate is $\propto P\lambda_{mfp}^2$ with $P$ as the laser power and $\lambda_{mfp}$ is the mean free path.[32,48] In addition to the electron-SPPs scattering, light induced heating can increase the temperature $(\delta T)$ [49] which can have reasonable influence on the device resistivity. Besides, electron-electron scattering rate also depends on the frequency $\omega$ of the irradiation and no longer it can be treated as independent of $\omega$ and hence the frequency dependence of $\tau$ should also be included to the total scattering rate in the illuminated condition.[50] An overall quadratic frequency dependence has been observed for the electron scattering rate on light illumination.[51,52] Thus in the presence of light with frequency $\omega$, the photo induced change in resistance, $\delta\rho$ can be expressed as (the detailed derivation is presented in the Supporting Information),

$$\delta\rho \propto \left(\omega^2 + \gamma_s P\lambda_{mfp}^2 \cdot \frac{\lambda_m}{L_{eff}}\right) \qquad (3)$$

where, $\gamma_s$ contains the information about the surface properties and the morphologically tuned distribution of SPPs along the interface between the metallic film and ambient air.[45]

However, morphology controlled effective electron density,[50] morphology controlled screening of plasma frequency under light illumination[47,53] and surface roughness controlled local light-induced electric field distribution have been reported earlier.[45,54] For the present study, we do not consider these effects since they need to be treated separately for individual structure type and dimension. $\lambda_m$ is the penetration depth. $L_{eff}$ relates to the effective dimension of nanostructured particles. The ratio of $\lambda_m$ and $L_{eff}$ represents the strength of the e-SPP interaction on the dimension of the morphological features.[32,55]

In a simplified manner, Equation (3) explains how the photo-induced resistivity depends on the frequency $\omega$ and the power intensity $P$ of the light irradiation. The other properties like surface roughness, morphology, particle dimension, penetration depth under irradiation, distribution of SPPs etc. are included in the second term through $\gamma_s$, $\lambda_m$, $L_{eff}$ in Equation (3). The summary of all the aforementioned scattering mechanisms are sketched in **Figure 7**.

From Equation (2), $I_{ph}$ is proportional to $\delta\rho$ and inversely proportional to the square of $\rho_{dark}$. First case: the observed photocurrent under light illumination with varying frequency and laser power for the same sample with no change in $\rho_{dark}$ can be understood from Equation (3). Figure 5(c) shows the wavelength dependence of photocurrent and the power dependent photocurrent for NIR light is shown in Figure S3(c)&(d) for a zone-III sample with morphologically varying nano and micro-structures. We observe that the amplitude of the photocurrent is about double for 532 nm laser than the same with 1064 nm laser. Note that the power difference is not much between 532 nm and 1064 nm lasers and hence the enhanced photocurrent for 532 nm light cannot be only due to the difference in power [Figure 5(c)]. Here the frequency dependent bulk scattering rate appeared in the first part of Equation (3) comes into play in addition to e-SPP interaction induced scattering contribution via the second term in Equation (3). And hence the first term contributes more in the visible range than the same in NIR range. The linear dependence of the photocurrent on the laser power in NIR range displayed in Figure S3(d) is justified by the second term in Equation (3) with fixed $\omega$.

Second case: the light source is fixed but variations in samples which are selected from three different zones as shown in Figure 6 with alterations in film thickness, morphology, surface roughness, etc. We have displayed only the result for halogen light offering the strongest response among all three light sources used in this study. We have studied the same with other two light sources also and a similar behaviour is observed for the 532 nm visible light [Figure S4]. For NIR 1064 nm laser, only the zone-III sample showed response [bottom panel of Figure 5(c), Figure S3 (c) & (d)] while the samples from other two zones did not exhibit any light-induced response. From Equation (5b), (7), and (9) in the Supporting Material, the surface properties are important in controlling $\delta\rho$ and $\rho_{dark}$. Experimentally, the surface morphology, embedded with structures like nanoparticles, nano and micro crystals and crystalline micro flowers with thickness variations from 30 nm to 150 nm as seen by the AFM, for zone-III samples is strikingly different than the same for other two zones as shown in Figure 6(a)-(f). Hence a dominant contribution from the surface to the photo-induced resistance $\delta\rho$ is expected for the zone-III samples with NIR exposure. In the visible range, zone-II samples respond to light also but with relatively lower photocurrent compared to that observed for zone-III samples. Here, the frequency dependent bulk scattering, i.e. the 1$^{st}$ term in Equation (3) might contribute the major part for the observed photocurrent. Finally, zone-I samples respond none of the studied three light sources. Due to much thinner films in zone-I where the film thickness is comparable to the electron mean free path,[56] the dark state resistivity $\rho_{dark}$ is expected to be higher[56,57] than that for the other two relatively thicker zones, hence the inverse square dependence on $\rho_{dark}$ dominates over the linear dependence on $\delta\rho$ for the photocurrent. Recently, it has also been shown for TiN films with thickness > 140 nm, resistivity increases inversely with thickness.[58] Therefore from Equation (2), the overall photocurrent is negligible with respect to the experimental noise level.

## 4. Conclusion

Synthesis and optoelectronic properties of ternary Ti-Si-N based composite thin films have been studied in this work. The main results and achievements are summarized as follows:

- A novel nitridation technique has been established in this study to prepare thin films of Ti-Si-N based transition metal nitride and silicide composite material with morphologically varying nano and micro crystalline structures like nanoparticles, faceted micro crystals, and self-assembled crystalline micro flowers etc.
- The novelty of the process for the nitridation relies solely on the substrate [Si (100) with 80 nm thick Si3N4 insulating spacer layer on top]. A smart execution of the reactive nature of Ti to $Si_3N_4$ at high temperature is the key factor for the demonstrated synthesis process.
- The coexistence of nitride and silicide phases of the metal in the composite film with a dominant contribution from the cubic TiN phase is revealed by HRXRD and HRTEM analysis.
- The composite films respond to light illumination and the photo-response depends on the surface morphology. The morphology dependent photo-response for the composite films leads to their suitability towards optoelectronic and plasmonic based applications.
- The observed experimental findings are explained unambiguously with a proposed model based on light induced scattering mechanisms for conduction electrons in the visible and NIR range. The model presented here is a simple but generic model which can be beneficial for understanding similar type of photo-response of other metallic low dimensional systems.

**Supporting Information**

Supporting Information is available from the Wiley Online Library or from the author.

**Acknowledgements**

The technical help for AFM, HRXRD, HRTEM, FESEM imaging along with EDS characterization and thickness optimization using the central facilities at CSIR-NPL are highly acknowledged. We are thankful to Mr. M. B. Chhetri and Ms. Shafaq Kazim for their assistance in the lab. S.Y. acknowledges financial support from UGC-JRF for providing junior research fellowship and B.G. acknowledges financial support from UGC-RGNF for providing senior research fellowship.

Received: ((will be filled in by the editorial staff))

Revised: ((will be filled in by the editorial staff))

Published online: ((will be filled in by the editorial staff))

**Conflict of Interest**

The authors declare no conflict of interest.

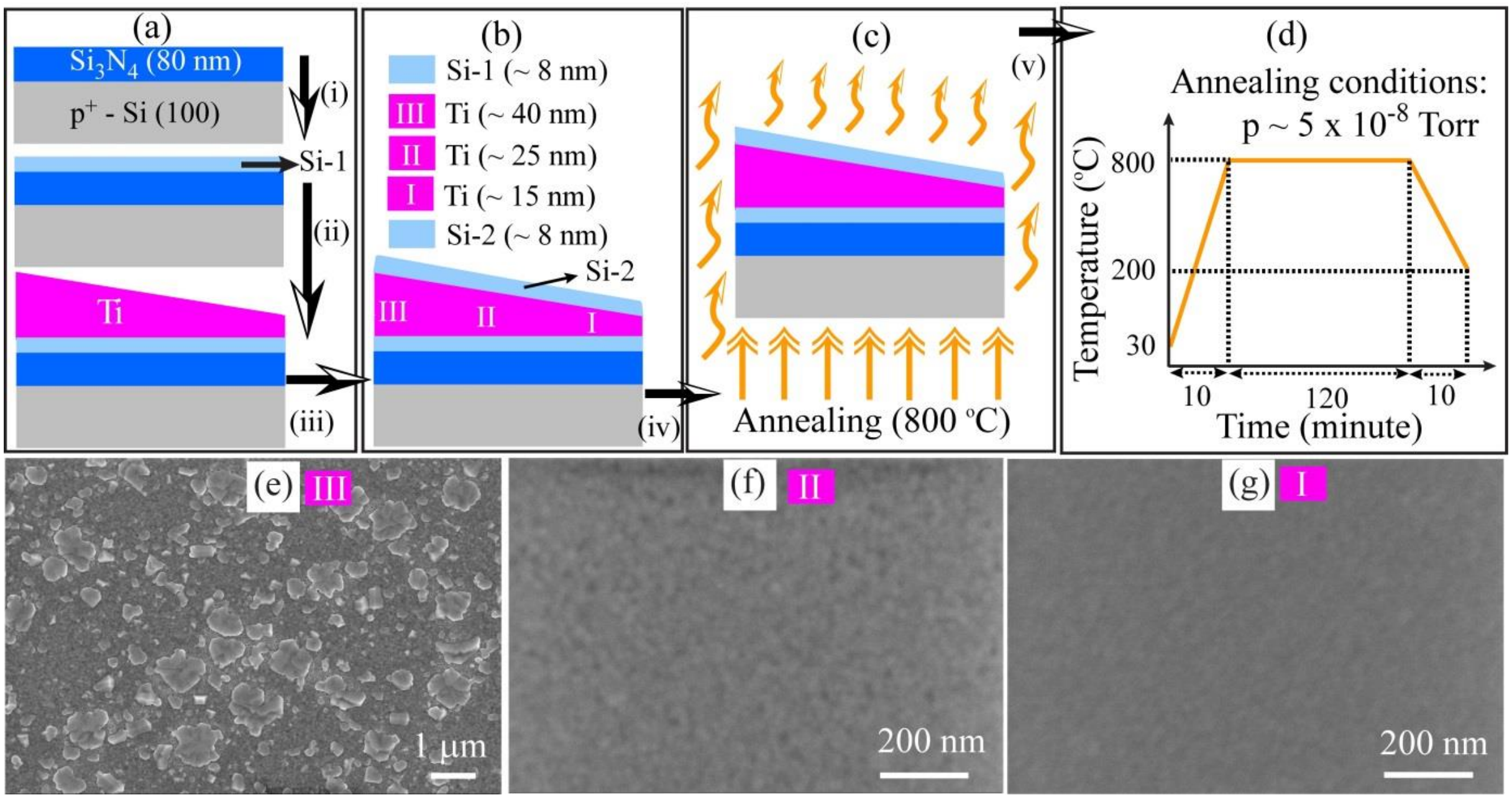

**Figure 1.** (a)-(d) Schematic presentation of the synthesis process for the fabrication of crystalline TiSiN nano and micro structures. Depending on the Ti thickness three different zones are marked in (b) by uppercase Roman numerals. The black arrows indicate the steps followed in the process. (e)-(g) FESEM images obtained from zones III, II and I respectively.

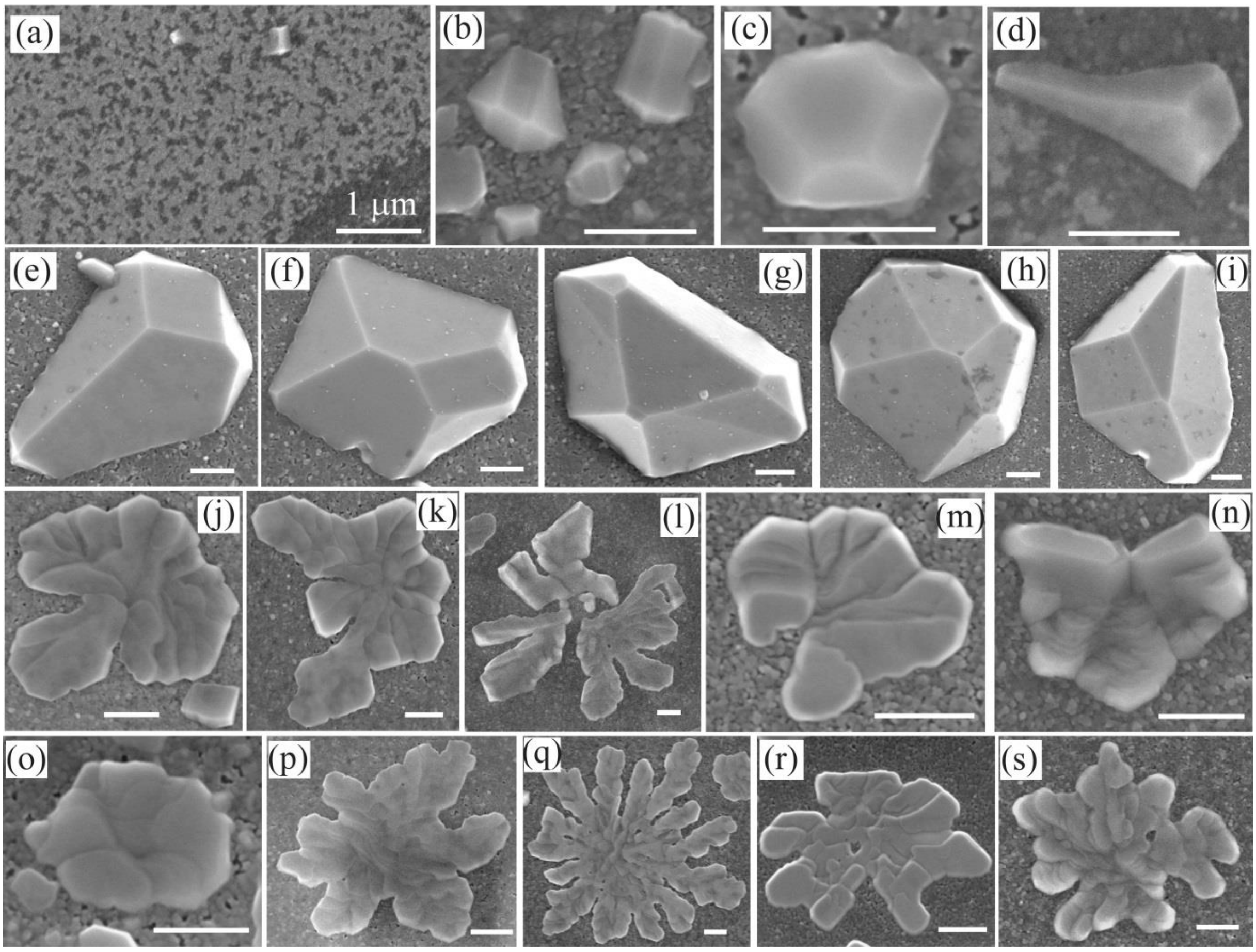

**Figure 2.** Categorical representation of morphologically different nano and microstructures collected from zone-III. (a) Appearance of a granular base layer constituted with spherical nanoparticles. (b) - (i) Collection of faceted microcrystals with different dimensions. Last two rows [(j) to (s)]: Self-assembled flower like structures. The second last row [(j) – (n)] represents the incomplete assembly while the last row [(o) – (s)] displays the complete formation of the rounded flowers. The scale bars represent 400 nm except (a).

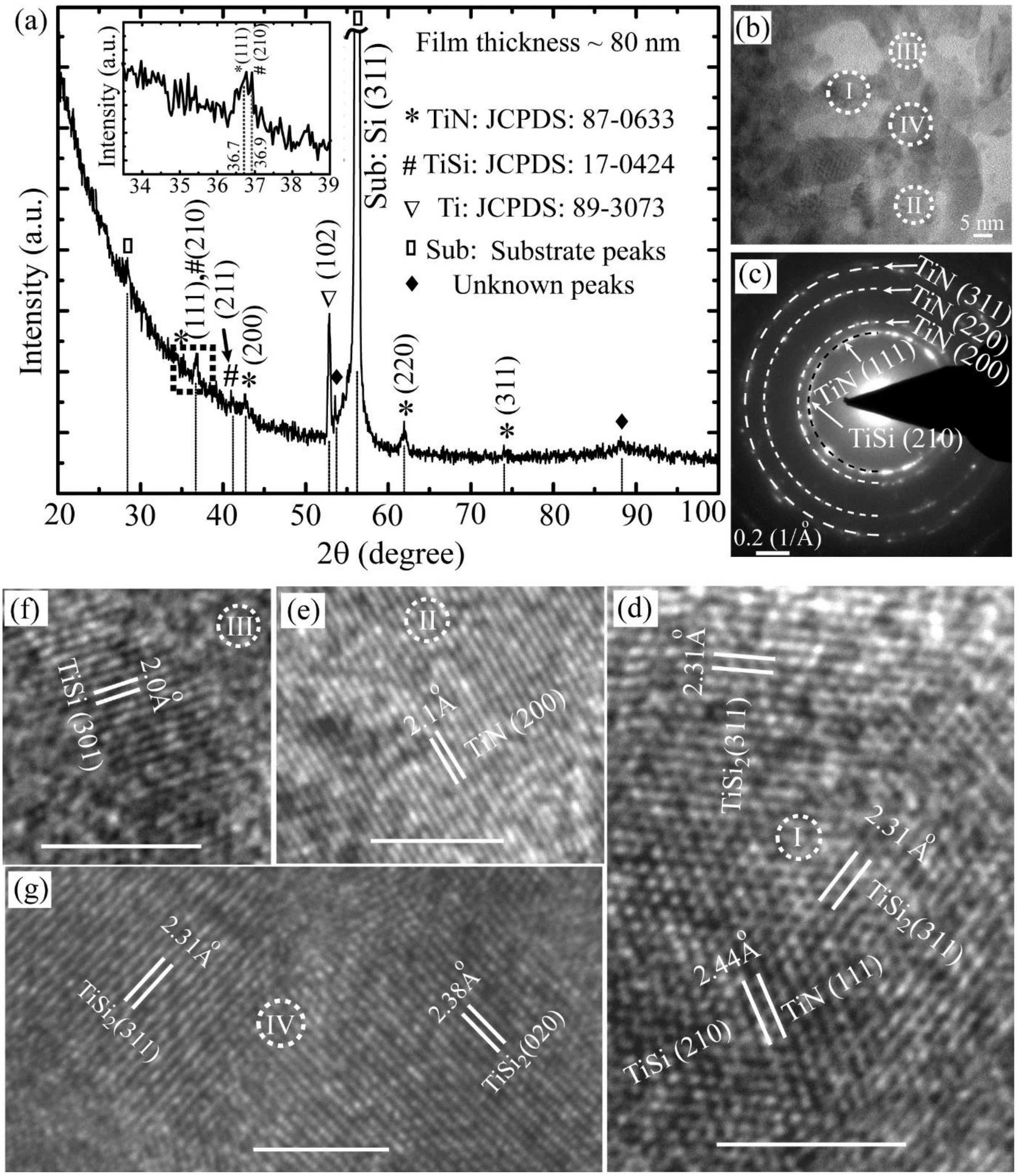

**Figure 3.** Structural characterization of the nanostructured particles. (a) The XRD pattern on zone (III) of a Ti-Si-N thin film. The dotted rectangular portion is zoomed in the inset which presents two distinct peaks related to TiN (111) and TiSi (210) phases. (b)-(g) TEM characterization of the nano-granular thin film mat. (b) Bright-field TEM micrograph of a part of the mat showing the constituent crystalline particles. Corresponding SAED pattern is shown in (c). (d)-(g) The HRTEM images of individual particles. The related particles/regions are marked with white dotted circles in (b). The scale bar is 3 nm.

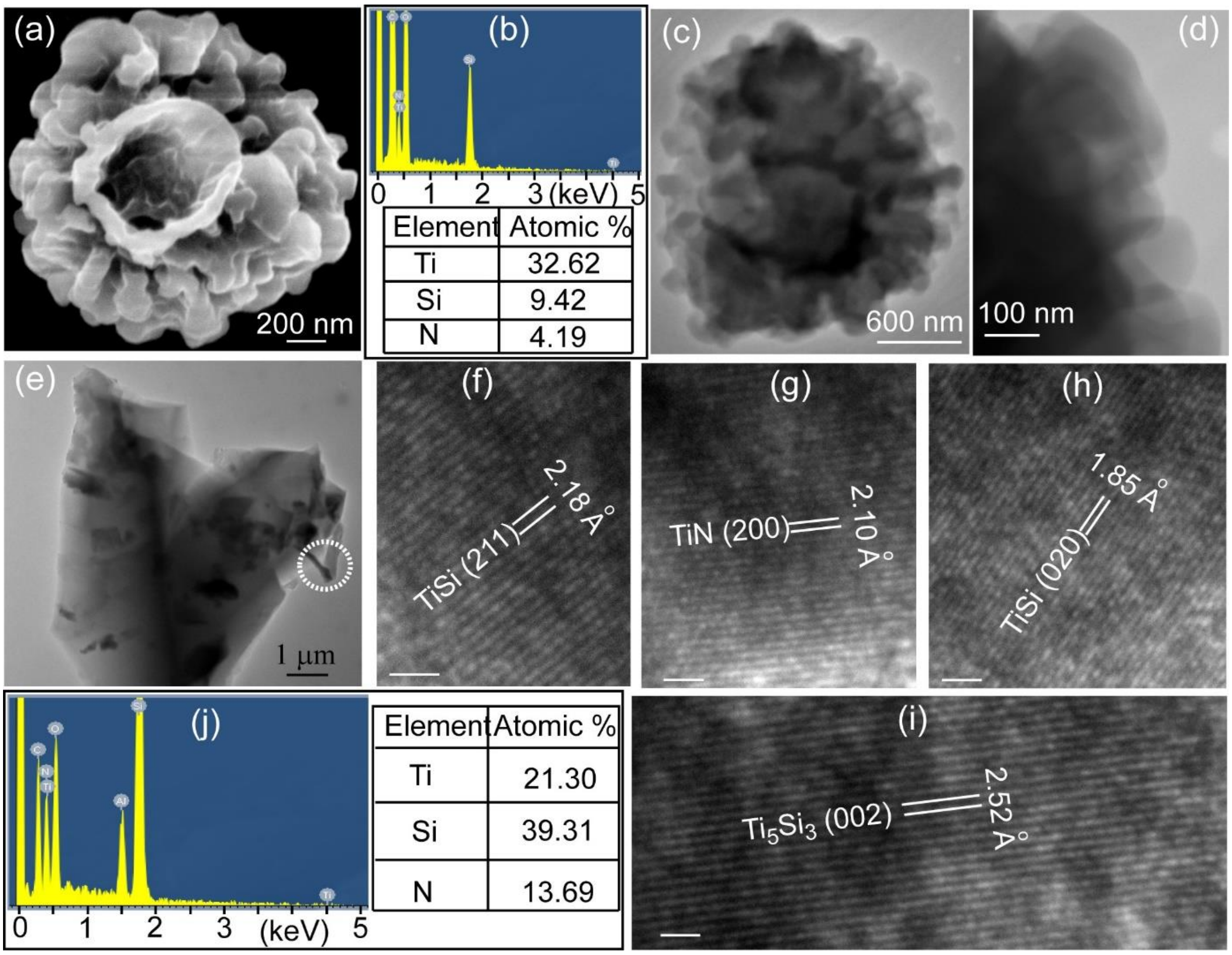

**Figure 4.** Structural characterization of micro structures. The EDS and TEM characterization of an individual flower (top row) and a crystalline flake (bottom two rows). (a) The FESEM image, (b) EDS elemental analysis and (c) TEM image for the flower. (d) A selected part of the flower demonstrating the layering of crystalline planes. (e) TEM image and (j) EDS elemental analysis for a flake. (f)-(i): HRTEM images obtained from a selected relatively thinner part of the flake as highlighted in (e). The scale bar is 1 nm where it is not mentioned.

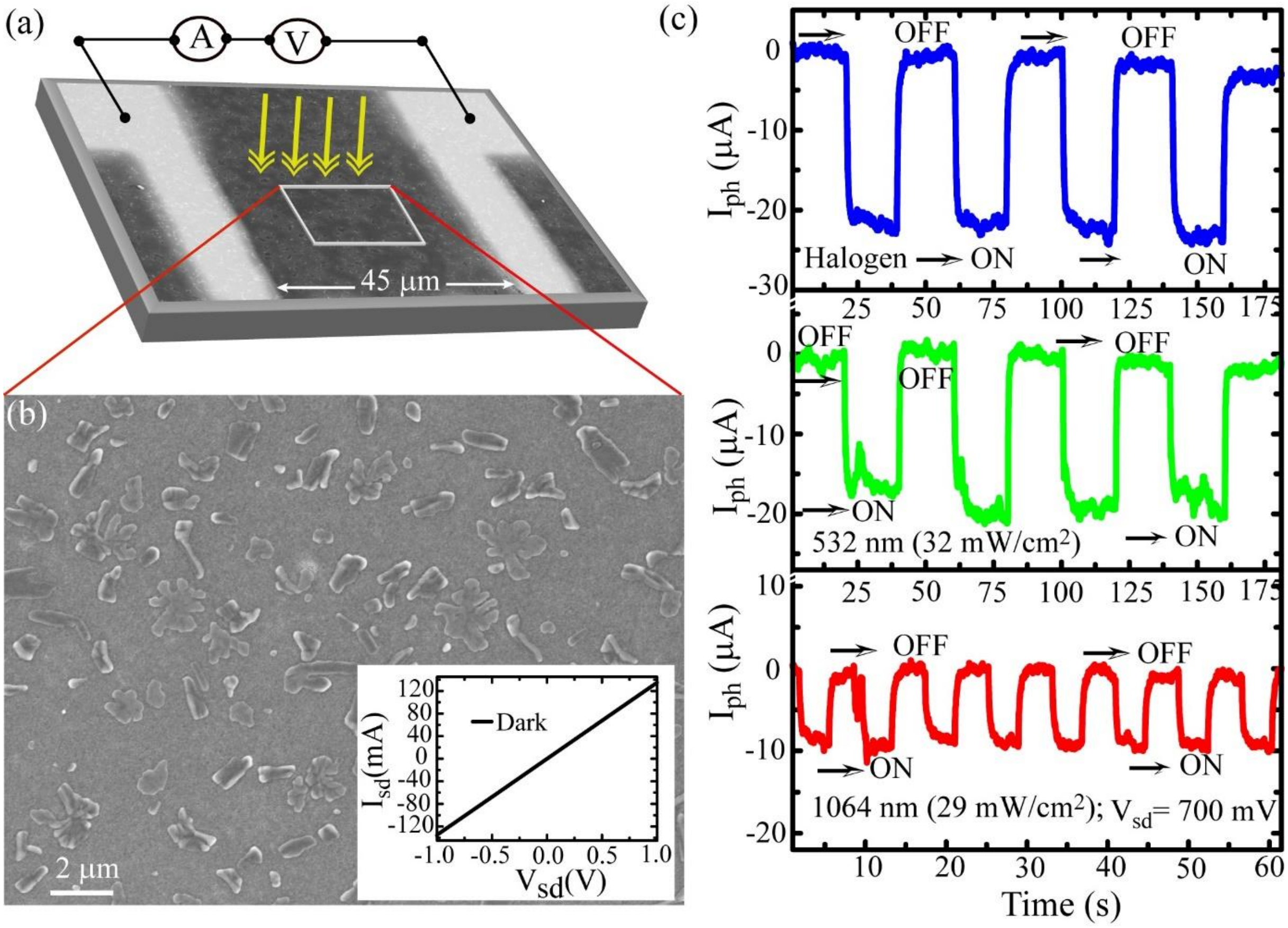

**Figure 5.** Anomalous photoconductivity (a) Schematic presentation of the transport measurement with a real device under optical illumination. (b) The magnified image of the surface morphology between the contact leads of the device as highlighted by the white box in (a). Inset: Current-voltage characteristic of the device in dark condition. (c) Time-resolved photo-response on the device current for repetitive light 'OFF' and 'ON' cycles. Light sources used here are halogen light (Top panel), 532 nm laser light in the visible region (Middle panel) and 1064 nm laser in NIR region (Bottom panel), respectively.

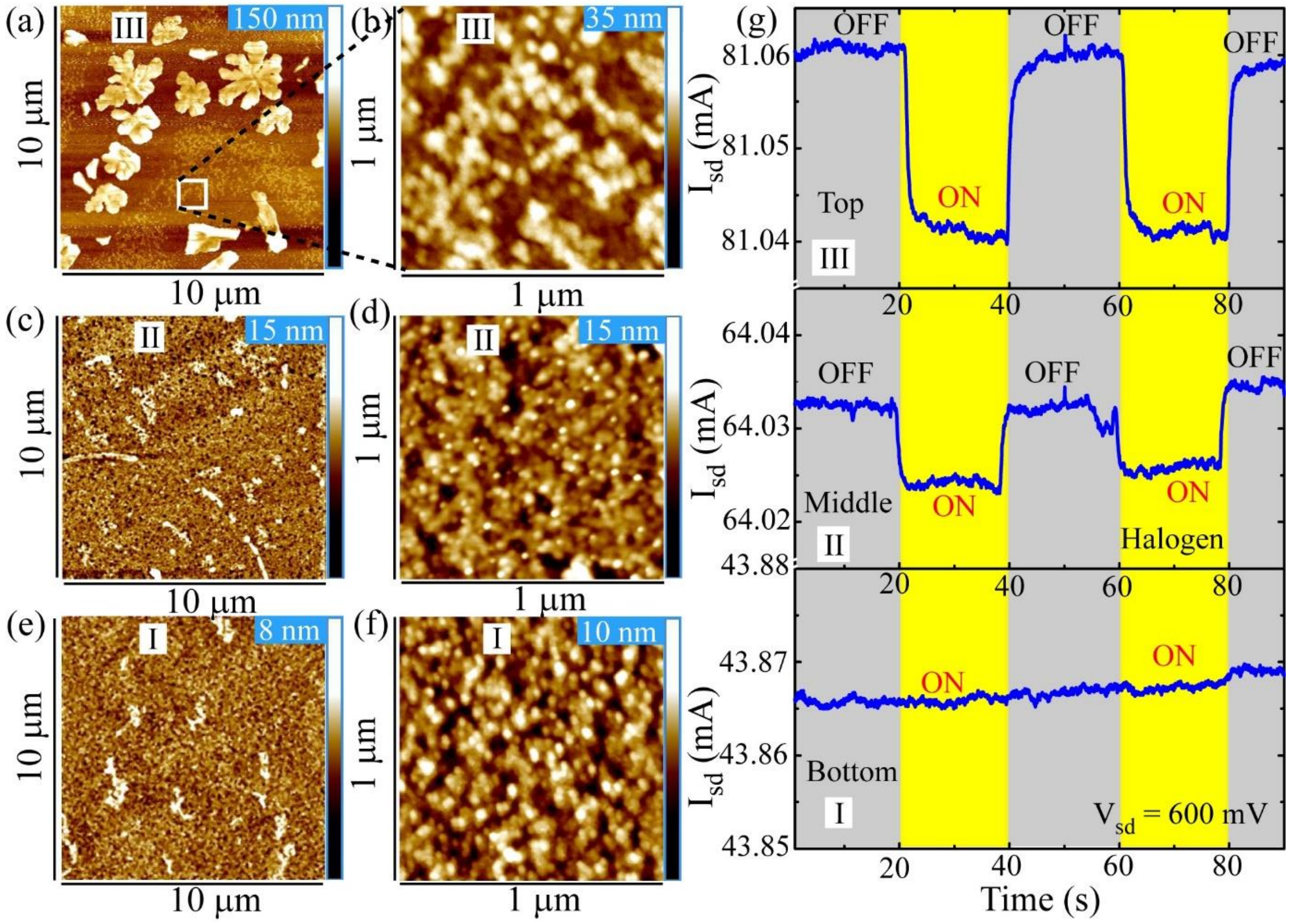

**Figure 6.** Morphology dependent photo-response. (a) - (f ) AFM topography images representing the morphology of three different zones. (b)Magnified image of a selected region, highlighted by the white square in (a). (c) & (e) relate the samples from zone-II and zone-I, respectively. Detailed surface morphologies for zone-II and zone-I samples are displayed using high resolution AFM images in (d) & (f), respectively. (g) The corresponding time-resolved fixed bias current for light 'OFF' and 'ON' states. The top, middle, and bottom panels represent the photo-response of the samples from zone-III, zone-II, and zone-I, respectively. Halogen light is used as the light source.

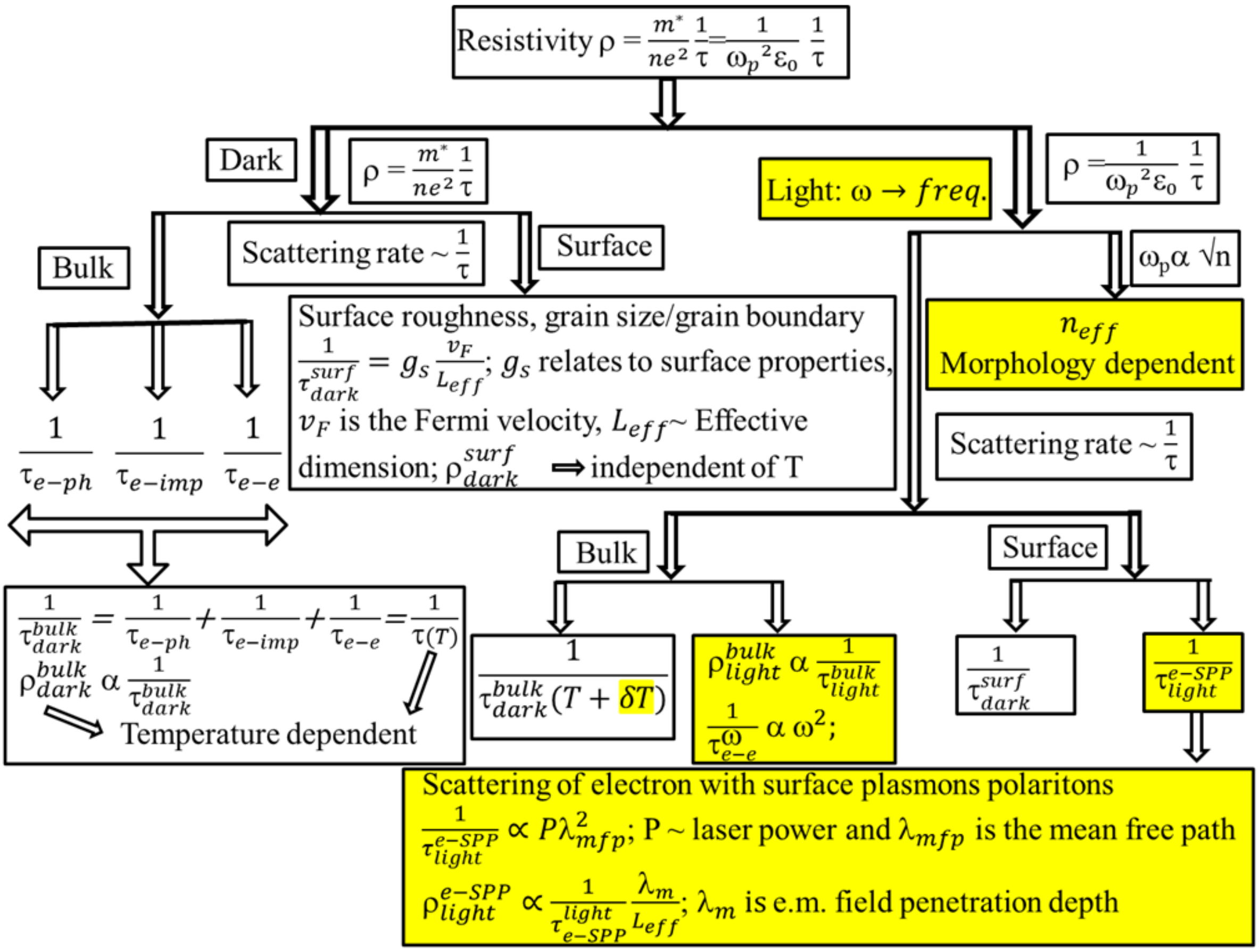

**Figure 7.** Summary of all the possible scattering mechanisms undergone by the conduction electron. The yellow shaded parts indicate the scattering under the light illumination.

# Supporting Information

## Substrate Mediated Synthesis of Ti-Si-N Nano-and-Micro Structures for Optoelectronic Applications

*Sachin Yadav, Alka Sharma, Bikash Gajar, Mandeep Kaur, Dinesh Singh, Sandeep Singh, Kamlesh Kumar Maurya, Sudhir Husale, Vijay Narain Ojha and Sangeeta Sahoo**

**Contents in the Supporting Information:-**

1. Comparison of XRD data on two different substrates.
2. Full EDS spectra showing all the elements present in the Ti-Si-N flake and the flower shown in Figure 3 in the main manuscript.
3. Bias dependent photocurrent for visible light and power dependent photocurrent measurement for NIR light for zone-III samples.
4. Comparison of zone-III and zone-II samples with respect to their photoresponse for 532 nm laser or visible light.
5. Mathematical derivation part from Drude's formula for dc resistivity under light illumination.
6. Table S1: Collection of all the samples into a Tabular form demonstrating the reproducibility for Ti-Si-N structures and their optoelectronic responses.

**(1) Comparison of XRD data on two different substrates:**

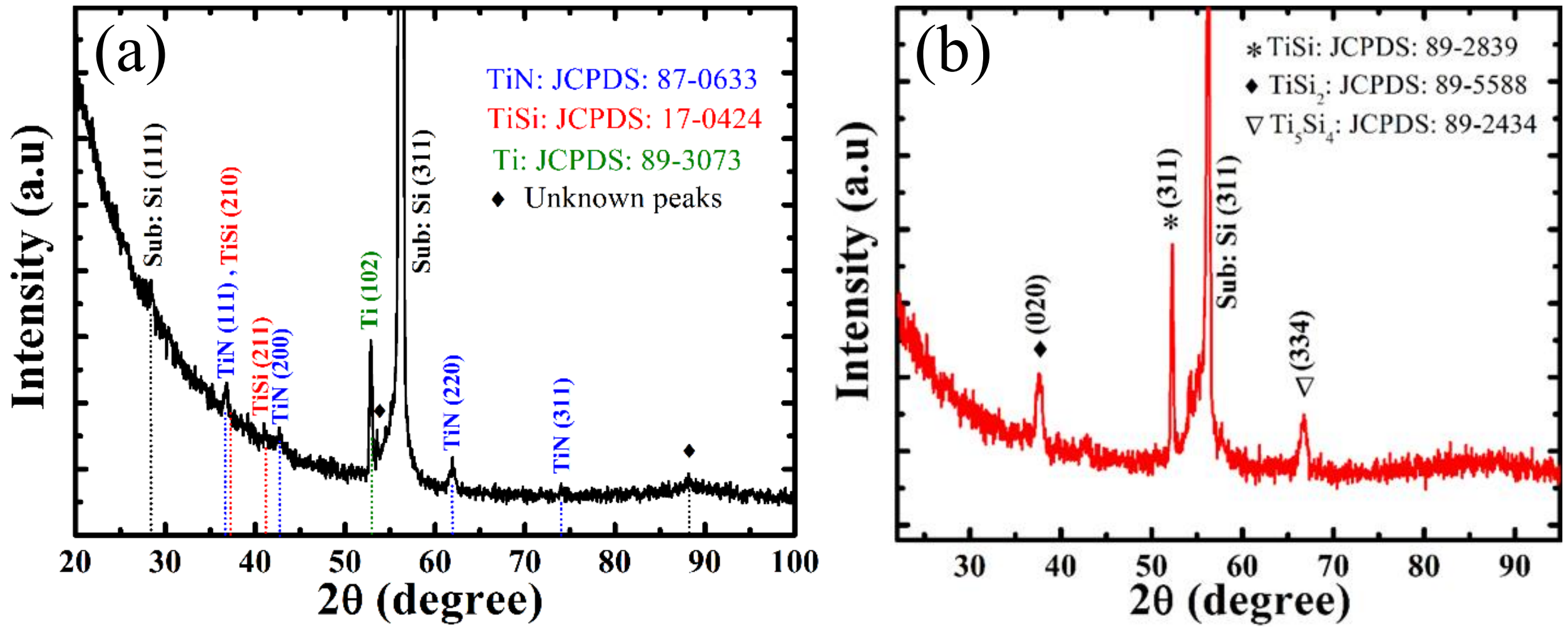

Figure S1. Comparison of XRD data measured on two different samples grown in same batch with same growth parameters but on two different substrates. (a) XRD pattern showing the co-existence of titanium nitride phases with the titanium silicide phases on $Si_3N_4$ /Si substrate. The same is presented in Figure 3(a) in the main manuscript. (b) XRD data of a sample grown in the same batch as that shown in (a) on $SiO_2$/Si substrate. The existence of various titanium silicide phases are evident.

Figure S1(a) shows the XRD spectrum of a Ti-Si-N based nanostructured composite film (CNF) sample which is also shown Figure 3(a) in the main manuscript. As it is mentioned in the main manuscript that the CNF is majorly comprised with cubic titanium nitride phases but the same also indicates the presence of titanium silicide phases. However, some of the nitride and silicide phases are very closely spaced in 2θ value and therefore it is difficult to visualize both the phases in the presented XRD data. In order to get the confirmation for the existence of titanium silicide phases, we have performed XRD characterization on a sample grown in the same batch but on $SiO_2$/Si substrate having no source of nitrogen to form titanium nitride phases. While growing the samples, both the substrates were kept under the same growth parameters and adjacent to each other on the sample holder so that the thickness and the heat profile remain almost unaltered. The XRD pattern for the samples grown on $SiO_2$/Si substrate is shown in Figure S1(b). The presence of various silicide phases are evident from the XRD data shown in Figure S2(b). Depending on the substrate type the reaction of Ti with nitrogen and/or silicon may have preferential growth but the comparison in the XRD characterization presented in Figure S1 clearly shows that the silicide phases cannot be ignored for the Si3N4/Si substrate even though TiN takes a major role in the presented CNFs.

**(2) Full EDS spectra showing all the elements present in the Ti-Si-N flake and the flower shown in Figure 4 in the main manuscript.**

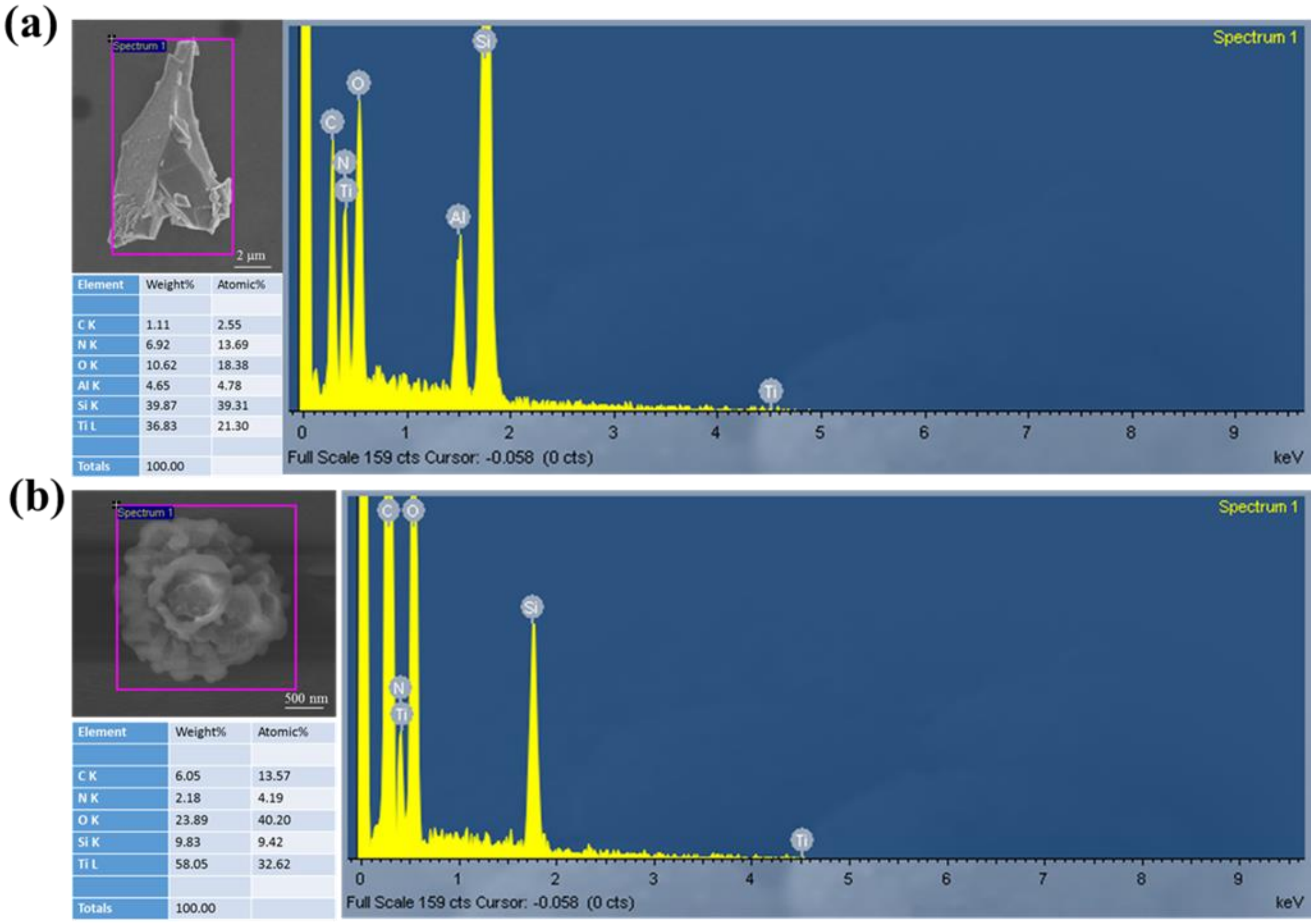

Figure S2. EDS spectra of a Ti-Si-N (a) flake and (b) flower

**(3) Bias dependent photocurrent for visible light and power dependent photocurrent measurement for NIR light for zone-III samples:**

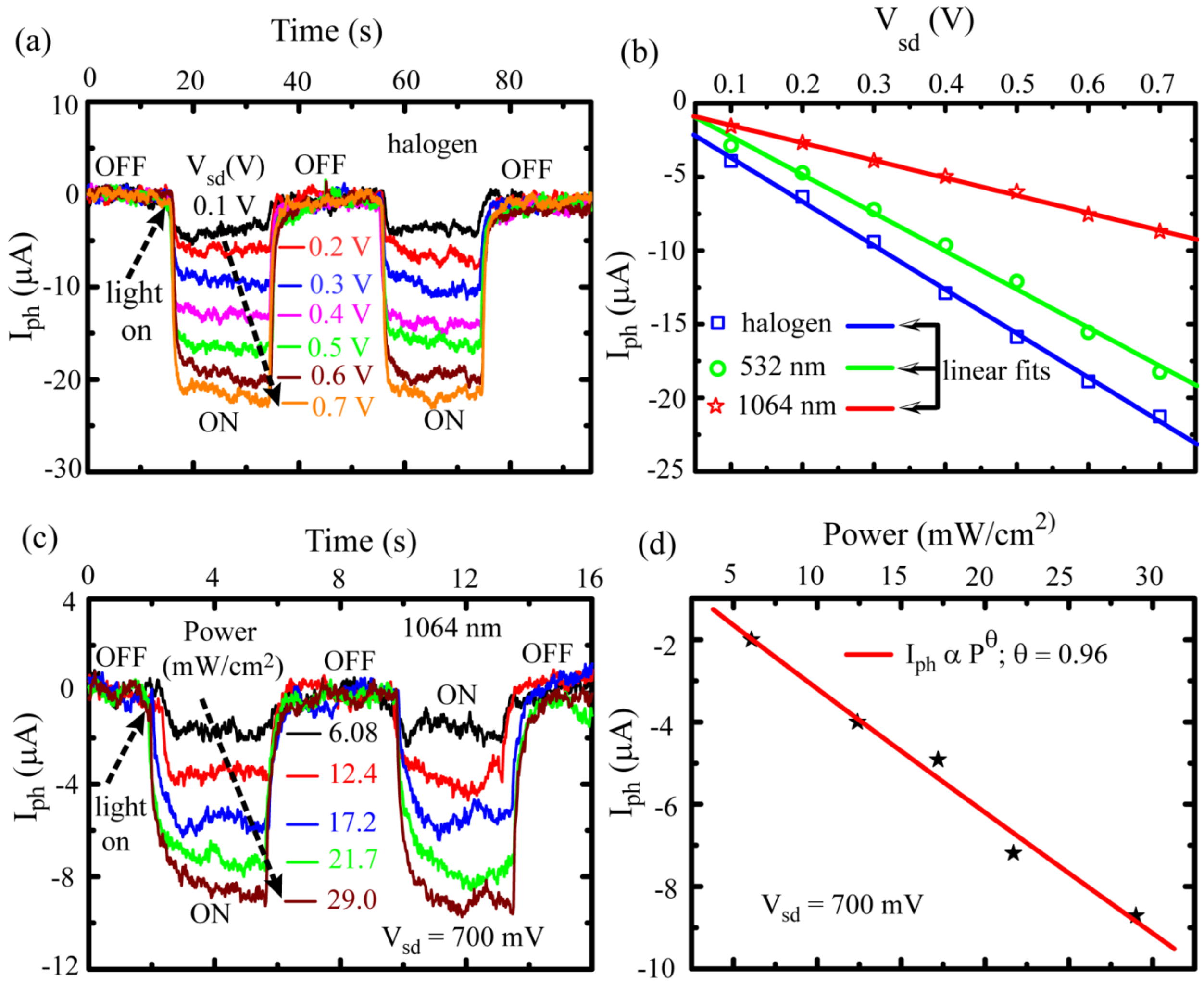

Figure S3. (a) Modulation of photo current ($I_{ph}$) with bias voltage for consecutive two cycles of (halogen) light 'OFF' and 'ON' states. (b) The bias dependent photocurrent, extracted from individual light 'OFF' and 'ON measurements as shown in (a), for halogen (the blue squares), 532 nm laser (the green open circles) and 1064 nm laser (the red stars). The solid lines are the fits with a sub-linear dependence of photo current to the bias voltage. (c) A set of fixed bias photo-current with varying power intensity of the 1064 nm laser for two consecutive light 'OFF' and 'ON' state. (d) Power dependent photo-current (black stars) for a fixed bias as collected from the individual curves in (c). The red solid curve is a power law fit.

For gaining an insight into the light sensing behaviour of the fabricated Ti-Si-N based devices we study the effect of conventional parameters like bias voltage, power of the light sources and the surface morphology on the measured photocurrent. The morphology dependence of the photo-response is presented in Figure 6 in the manuscript and here we show the bias voltage dependence and laser power dependence of the photocurrent for zone-III samples only. Figure S3 represents the bias dependent photocurrent for all the three mentioned light sources and the power dependent photocurrent in the NIR region with 1064 nm laser light. A representative set of time resolved photocurrent differing in bias voltage for two consecutive cycles of light 'OFF' and 'ON' states with halogen light is shown in Figure S3(a). The amplitude of the photocurrent evidently modulates with the bias voltage. We have extracted individual photocurrent values for each bias voltage from the time resolved photocurrent measurements as shown in Figure S3(a). The corresponding photocurrent values are plotted with the applied bias voltage in Figure S3(b). The scattering points represent the extracted experimental points and the solid lines are the linear fits. The linear dependence of the photocurrent on bias voltage confirms the suitability of the morphologically enriched Ti-Si-N nanostructured composite film (NCF) for the photodetector based applications [1, 2]. Figure S3(c) & (d) exhibit the variation of the photocurrent with the power intensities for 1064 nm laser. Similar to Figure S3(a), we have displayed a set of time dependent photocurrent varying with the power of the laser light for two consecutive 'OFF' and 'ON' cycles in Figure S3(c). The applied bias voltage was fixed at 700 mV. The amplitude of the photocurrent increases with the power. In Figure S3(d), we illustrate the dependence of photocurrent on laser power. The black stars represent the experimental values and the red solid curve is a power law fit using the conventional equation, $I_{ph} \propto P^{\theta}$ with $P$ as the power of the light and $\theta$ being the exponent. The best fit shown in Figure S3(d) offers almost a linear dependence for the photocurrent on the laser power with the exponent as θ ~0.96. This is again a characteristic feature which should be maintained for a good photodetector [3].

Besides, the linear dependence of photocurrent on the power is the indication of a very low density of trapped states present at the Fermi surface [4,5].

Finally, we have shown that Ti-Si-N NCFs embedded with nano-grains and self-assembled microcrystalline structures respond to the light in the visible-to-NIR spectral range in a similar fashion how a photodetector performs [6,7] except for the anomalous nature of the photocurrent [5,8]. Note, the performance level and its comparison with the established photodetectors are not the main subject of this study and we believe that a rigorous optimization and control over the growth parameters are needed to achieve the best performance level. Here, mainly we emphasize that Ti-Si-N based composite material can perform interesting optoelectronic characteristics too in addition to their proven excellence in other fields.

**(4) Comparison of zone-III and zone-II samples with respect to their photoresponse for 532 nm laser**

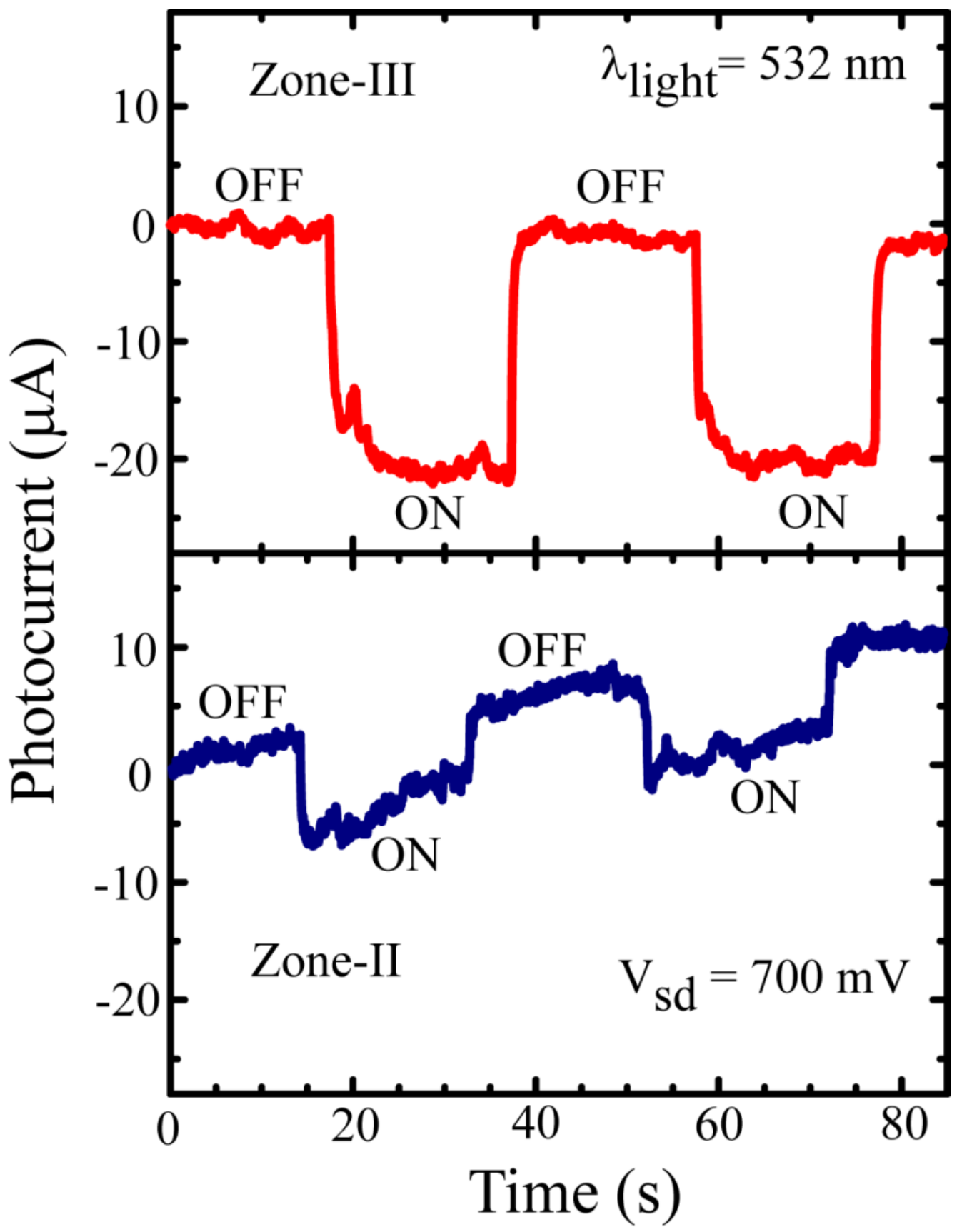

Figure S4. Time dependent photocurrent measurements for zone-III (top panel) & zone II (bottom panel) samples with 532 nm laser in the visible range. The bias voltage was fixed at $V_{ds}$ =700 mV. Corresponding morphological changes are shown in Figure 6 in the main manuscript.

Figure S4 demonstrates the time-resolved photocurrent measurement under 532 nm visible laser light for zone-III (top panel) and zone-II (bottom panel) samples, respectively. The current was measured with a fixed bias, Vsd = 700 mV. For comparison, we have used the same scale to present the amplitude of photocurrent measured from samples of both zones. As we have observed for the white light presented in Figure 6, the amplitude of the photocurrent for zone-III sample is almost double of the same from zone-II sample. Consequently, the photoresponse is much stronger and prominent for zone III samples than that for the zone-II samples for 532 nm laser. This is quite understandable from the proposed model with the frequency and morphology induced surface plasmon polariton (SPPs) dependent scattering rates presented by eq. (12). For zone-III samples, the contribution from the e-SPPs interaction adds to the total scattering rate and hence a stronger response to light by increasing the device resistivity is observed. Whereas for zone-II samples, the varieties in the morphological structures on the surface are absent and hence the e-SPPs related surface contribution might not add significantly to the total electron scattering and hence the light induced resistivity change can only be due to the frequency dependent scattering from the bulk. Further, the total thickness of zone-II is less compared to that of zone-III samples and thickness dependent dark resistivity can influence the overall amplitude of the photocurrent as explained by eq. (2) in the manuscript. Note, unlike the behavior with halogen light shown in Figure 6, the bottom panel for zone-II samples under 532 nm laser shows a upward slope in the measured current with time whereas, for zone-III samples in the top panel shows a stable situation with respect to light ‘OFF’ & ‘ON’ cycles. The slope occurring in bottom panel could be due to heating effect at the interface between contact leads and the film since the power of the laser 532 nm is high compared to that of the white light. However, the effect of light on the measured current is evident by clear switching from a higher value to a lower value in response to the light ‘ON’ condition.

**(5) Mathematical derivation part from Drude's formula for dc resistivity under light illumination:**

In order to understand the origin of the observed anomalous photoconductivity controlled mainly by the surface morphology and the film thickness, here we calculate the dc resistivity in the dark and illumination conditions.

As the devices exhibit linear current-voltage characteristics [inset of Figure 5(b)] the device current can be expressed as:

$$I_{dark} = \frac{V_{sd}}{R_{dark}} \propto \frac{1}{\rho_{dark}} \text{ and}$$

$$I_{light} = I_{dark} + \delta I = I_{dark} + I_{ph}$$

$$= \frac{V_{sd}}{R_{light}} \propto \frac{1}{\rho_{light}}; \text{ with}$$

$$\frac{1}{\rho_{light}} = \frac{1}{\rho_{dark} + \delta\rho} = \frac{1}{\rho_{dark}\left(1 + \frac{\delta\rho}{\rho_{dark}}\right)}$$

$$= \frac{1}{\rho_{dark}}\left(1 + \frac{\delta\rho}{\rho_{dark}}\right)^{-1} \cong \frac{1}{\rho_{dark}}\left(1 - \frac{\delta\rho}{\rho_{dark}}\right)$$

for $\frac{\delta\rho}{\rho_{dark}} \ll 1$; higher order terms are ignored

Therefore,

$$I_{light} \propto \frac{1}{\rho_{dark}}\left(1 - \frac{\delta\rho}{\rho_{dark}}\right) \quad (1)$$

and

$$I_{ph} = \delta I \propto \frac{\delta\rho}{\rho_{dark}^2} \quad (2)$$

Where, *I*, *R*, $V_{sd}$, $\rho$ represent current, resistance, source-drain bias voltage, and resistivity, respectively. The subscript *dark* (*light*) represents the measured values under light '*OFF*' ('*ON*') state, respectively. $I_{ph}$ is the photocurrent.

From Equation (2) $I_{ph}$ depends on dark state resistivity ($\rho_{dark}$) and the photo-induced resistivity ($\delta\rho$). Thus for a clear understanding of the origin behind the observed NPC, we now consider the effect of surface morphology, thickness and the light illumination on the device resistivity. In general, $\rho_{dark}$ is constant for a device when only the effect of light is considered while keeping the device parameters like the film thickness, surface roughness etc.

unaltered. In this case, $I_{ph}$ depends only on $\delta\rho$. Now we address possible mechanisms controlling the device resistivity under the light exposure.

From Drude's free electron model dc resistivity, $\rho_{dc}$, can be expressed as,

$$\rho_{dc} = \frac{m_e^*}{n_e e^2 \tau} \propto \frac{1}{\tau} \qquad (3)$$

Where, $n_e$, $m_e^*$, $e$ are the free electron density, effective mass, and electronic charge, respectively. $\varepsilon_0$ is the permittivity of free space and $\tau$ represents the electron scattering time which depends mainly on the purity of the metal, charge state of the impurity and temperature. Considering $m_e^*$ and $n_e$ as constants in the dark state, the dc resistivity from Equation (3) depends mainly on the scattering time, $\tau$. Electron-phonon, electron-impurity and electron-electron scattering mechanisms are the dominant parts towards the bulk electron scattering in a metal. Surface electron scattering, controlled mainly by the surface roughness, grain size and grain boundary, also contribute significantly to the resistivity [9]. Hence the surface and the bulk contributions for the scattering rate can be added using Matthiessen's rule:

$$\frac{1}{\tau_{dark}} = \frac{1}{\tau_{dark}^{bulk}} + \frac{1}{\tau_{dark}^{surf}} \qquad (4)$$

$$\frac{1}{\tau_{dark}^{bulk}} = \frac{1}{\tau^{e-ph}} + \frac{1}{\tau^{e-imp}} + \frac{1}{\tau^{e-e}} = \frac{1}{\tau(T)} \qquad (5a)$$

and

$$\frac{1}{\tau_{dark}^{surf}} = g_s \frac{v_F}{L_{eff}} \qquad (5b)$$

Here, $\tau_{dark}^{bulk}$ and $\tau_{dark}^{surf}$ are bulk and surface contribution to the electron scattering time in dark, respectively. $g_s$, $v_F$, $L_{eff}$ in Equation (5b) relate to surface properties, Fermi velocity and the effective dimension of nanostructured particles, respectively[10]. For spherical particles, $L_{eff}$ represents the diameter (D) and for nanorod type of structures it can be represented by $(LD)^{1/2}$ with $L$ and $D$ being the length and the diameter of the rod, respectively [11]. Now, $\rho_{dc}$ can be constituted with its bulk ($\rho_{bulk}$) and surface ($\rho_{surf}$) contributions as:

$$\rho_{dc} = \rho = \rho^{bulk} + \rho^{surf} \qquad (6)$$

In dark state,

$$\rho_{dark} \propto \left(\frac{1}{\tau_{dark}^{bulk}} + \frac{1}{\tau_{dark}^{surf}}\right) \qquad (7)$$

Under the light illumination, surface plasmons (SPs), initiated and controlled by the surface roughness and morphology, takes a dominant role to control the device's transport response. Surface morphology controls the local electric field enhancement, [12-14] radiative damping [15] and non-radiative Landau damping [16, 17] for the SPs [18]. Further, coupled with the light excitation SPs can induce SPPs which contributes significantly to the total electron scattering rate while interacting with the conduction electron [19]. The electron-SPPs scattering rate can be expressed as, $\tau_{light}^{e-SPP} \propto P\lambda_{mfp}^2$ with $P$ as the laser power and $\lambda_{mfp}$ is the mean free path [19, 20]. In addition to the electron-SPPs scattering, light induced heating can increase the temperature ($\delta T$) [21] which can have reasonable influence on the device resistivity as shown in Equation (5a). Besides, electron-electron scattering rate also depends on the frequency $\omega$ of the irradiation and no longer it can be treated as independent of $\omega$ and hence the frequency dependence of $\tau$ should also be included to the total scattering rate in the illuminated condition [22]. An overall quadratic frequency dependence has been observed for the electron scattering rate on light illumination [23,24]. Thus in the presence of light with frequency $\omega$, the device resistivity can be written as,

$$\frac{1}{\tau_{light}} = \frac{1}{\tau_{light}^{bulk}} + \frac{1}{\tau_{light}^{surf}} \text{ ; now,}$$

$$\frac{1}{\tau_{light}^{bulk}} = \frac{1}{\tau^{e-ph}} + \frac{1}{\tau^{e-imp}} + \frac{1}{\tau^{e-e}} = \frac{1}{\tau(T',\ \omega)}$$

$$\propto \left(\frac{1}{\tau(T')} + \omega^2\right) \propto \left(\frac{1}{\tau(T+\delta T)} + \omega^2\right) \qquad (8)$$

and

$$\frac{1}{\tau_{light}^{surf}} = \frac{1}{\tau_{dark}^{surf}} + \frac{1}{\tau_{light}^{e-SPP}} = g_s\frac{v_F}{L_{eff}} + \gamma_s P\lambda_{mfp}^2 \qquad (9)$$

where, $\gamma_s$ contains the information about the surface properties and the morphologically tuned distribution of SPPs along the interface between the metallic film and ambient air [16]. Now the total resistivity in light 'ON' state can be expressed as:

$$\rho_{light} = \rho_{light}^{bulk} + \rho_{light}^{surf} \propto \left(\frac{1}{\tau_{light}^{bulk}} + \frac{1}{\tau_{light}^{surf}}\right) \propto \left[\frac{1}{\tau(T+\delta T)} + \omega^2 + g_s \frac{v_F}{L_{eff}} + \gamma_s P \lambda_{mfp}^2 \cdot \frac{\lambda_m}{L_{eff}}\right]$$

(10)

Here for simplicity, we have assumed the electronic mass and the electron density as constants under the illumination. However, morphology controlled effective electron density [22], morphology controlled screening of plasma frequency under light illumination [18,25] and surface roughness controlled local light-induced electric field distribution have been reported earlier [16,26]. For the present study we do not consider these effects since they need to be treated separately for individual structure type and dimension. $\lambda_m$ is the penetration depth. The ratio of $\lambda_m$ and $L_{eff}$ represents the strength of the e-SPP interaction on the dimension of the morphological features [19, 27]. Now, if we assume that $\delta T$ can have negligible contribution towards the room temperature resistivity, Equation (10) can be simplified as,

$$\rho_{light} = \rho_{dark} + \delta\rho \propto \begin{bmatrix} \left(\frac{1}{\tau(T)} + g_s \frac{v_F}{L_{eff}}\right) + \\ \left(\omega^2 + \gamma_s P \lambda_{mfp}^2 \cdot \frac{\lambda_m}{L_{eff}}\right) \end{bmatrix} \qquad (11)$$

From Equation (5a), (5b), and (7), the first two terms in the first parenthesis in Equation (11) represent the dark state resistivity, $\rho_{dark}$ and therefore the second parenthesis with the last two terms in Equation (11) relates the photo induced change in resistance, $\delta\rho$. Hence,

$$\delta\rho \propto \left(\omega^2 + \gamma_s P \lambda_{mfp}^2 \cdot \frac{\lambda_m}{L_{eff}}\right) \qquad (12)$$

In a simplified manner, Equation (12) explains how the photo-induced resistivity depends on the frequency $\omega$ and the power intensity $P$ of the light irradiation. The other properties like surface roughness, morphology, particle dimension, penetration depth under irradiation, distribution of SPPs etc. are included in the second term through $\gamma_s$, $\lambda_m$, $L_{eff}$ in Equation (12).

**(6) Table S1: Collection of all the samples into a Tabular form demonstrating the reproducibility for TiSiN structures and their optoelectronic responses.**

| Sample No. | Batch No. | Si Thick-ness | Ti Thick--ness | Si Thick-ness | Annea-ling Temp ($^0$C) | Annea-ling Time | Morpho-logy | Optoelectronic measurements results |
|---|---|---|---|---|---|---|---|---|
| **S#1** | B-188 | 20 nm | 80 nm | 6 nm | 760 | 55 min | Zone-I | No Photoresponse |
| **S#2** | B-189 | 6 nm | 104 nm | 8 nm | 800 | 90 min | Zone-III | Negative Photoresponse |
| **S#3** | B-191 | 8 nm | 80 nm | 8 nm | 780 | 120 min | Zone-III | Negative Photoresponse |
| **S#4** | B-192 | 4 nm | 60 nm | 8 nm | 800 | 120 min | Zone-II | No measurements were performed |
| **S#5** | B-194 | 8 nm | 80 nm | 8 nm | 800 | 120 min | Zone-III | Negative Photoresponse |
| **S#6** | B-205 | 8 nm | 80 nm | 8 nm | 800 | 120 min | Zone-III | Negative Photoresponse |
| **S#7** | B-207 | Nil | 104 nm | Nil | 800 | 120 min | Zone-I | No Photoresponse |
| **S#8** | B-208 | 40 nm | 70 nm | 6 nm | 800 | 120 nm | Zone-I | No Photoresponse |
| **S#9** | B-209 | 8 nm | 80 nm | 8 nm | 800 | 120 min | Zone-III | Negative Photoresponse |
| **S#10** | B-213 | 8 nm | 80 nm | 8 nm | 800 | 120 nm | Zone-III | No measurements were performed |
| **S#11** | B-215 | 8 nm | 80 nm | 8 nm | 800 | 120 min | Zone-III | Negative Photoresponse |
| **S#12** | B-215 | 8 nm | 50 nm | 8 nm | 800 | 120 min | Zone-II | Negative Photoresponse |
| **S#13** | B-215 | 8 nm | 30 nm | 8 nm | 800 | 120 min | Zone-I | No Photoresponse |
| **S#14** | B-230 | 8 nm | 70 nm | 8 nm | 800 | 120 min | Zone-III | Negative Photoresponse |
| **S#15** | B-230 | 8 nm | 40 nm | 8 nm | 800 | 120 min | Zone-II/ Zone-I | No Photoresponse |
| **S#16** | B-244 | 8 nm | 80 nm | 8 nm | 800 | 120 min | Zone-III | No measurements were performed |